\documentstyle[aps,preprint]{revtex}
\begin{document}
 
\begin{center}
{\Large \bf  Multiplicative processes and power laws}
 
\vskip .17in
Didier Sornette$^{1,2}$
 
{\it $^1$ Laboratoire de Physique de la Mati\`ere Condens\'ee, CNRS
UMR6632\\ Universit\'e des Sciences, B.P. 70, Parc Valrose, 06108 Nice Cedex
2,
France \\
 
$^2$ Department of Earth and Space Science\\ and Institute of Geophysics
and
Planetary Physics\\ University of California, Los Angeles, California
90095, USA\\
 
}
\end{center}
 \vskip 3cm
\noindent
{\Large \bf Abstract} Takayasu et al. 
\cite{Taka} have revisited the question of stochastic processes with multiplicative noise, which
have been studied in several different contexts over the past decades.
We focus on the regime, found for a generic set of control parameters, in which
stochastic processes with multiplicative noise produce intermittency of a special
kind, characterized by a power law probability density distribution. 
We briefly explain the physical mechanism leading to a power law pdf and
provide a list of references for these results
 dating back from a quarter of century. We explain
how the formulation in terms of the characteristic function developed by 
Takayasu et al.  \cite{Taka}  can be extended to
exponents $\mu >2$, which explains the ``reason of the lucky coincidence''.
The multidimensional generalization of (\ref{eq1}) and the available results are 
briefly summarized. The discovery of stretched exponential tails in the presence of the cut-off
introduced in \cite{Taka} is explained theoretically. We end by briefly listing applications. 

\pagebreak



\section{Stochastic multiplicative processes repelled from the origin}

Takayasu et al. \cite{Taka} recently studied the discrete stochastic equation
\begin{equation}
x(t+1) = b(t) x(t) + f(t) ~,
\label{eq1}
\end{equation}
as a generic model for generating power law pdf (probability density function).
Eq.(\ref{eq1}) defines a stationary process if $\langle \ln b(t) \rangle < 0$.

In order to get a power law pdf, $b(t)$ must sometimes take values larger than one,
corresponding to intermittent amplifications. 
This is not enough: the presence of the additive term $f(t)$ (which can
be constant or stochastic) is needed to ensure a ``reinjection'' to finite values, 
susceptible to the intermittent amplifications. It was thus shown \cite{Sorcont}
that (\ref{eq1}) is only one among many convergent ($\langle \ln b(t) \rangle < 0$)
multiplicative processes with 
repulsion from the origin (due to the $f(t)$ term in (\ref{eq1})) of the form
\begin{equation}
x(t+1) = e^{F(x(t), \{b(t), f(t),...\})} ~b(t) ~x(t)~,
\end{equation}
 such that $F \to 0$ for 
large $x(t)$ (leading to a pure multiplicative process for large $x(t)$)
 and $F \to \infty$ for $x(t) \to 0$ (repulsion from the origin). 
$F$ must obey some additional constraint
such a monotonicity which ensures that no measure is concentrated over a finite interval.
All these processes share the
same power law pdf 
\begin{equation}
P(x) = C x^{-1-\mu}
\label{er}
\end{equation}
 for large $x$ with $\mu$ solution
of 
\begin{equation}
\langle b(t)^{\mu} \rangle = 1~.
\end{equation}

The fundamental reason for the existence of the powerlaw pdf (\ref{er}) is that
 $\ln x(t)$ undergoes a random walk with
drift to the left and which is repelled from $-\infty$. A simple Boltzmann argument \cite{Sorcont}
shows that the stationary
concentration profile is exponential, leading to the power law pdf in the $x(t)$
variable.

These results were proved for the process (\ref{eq1}) by Kesten \cite{Kesten}
using renewal theory and was then revisited by several authors
in the differing contexts of ARCH processes in econometry \cite{Haan} and 1D random-field Ising
models \cite{Calan} using Mellin transforms, and more recently using 
extremal properties of the $G -${\it
harmonic} functions on non-compact groups \cite{Solomon} and the Wiener-Hopf 
technique \cite{Sorcont}. Many other results are available, for instance
concerning the extremes of the 
 process $x(t)$ \cite{emmbre} which shows that $x(t)$ have similar extremal
 properties as a sequence of iid random variables with the
 same pdf. The 
 subset of times $1 \leq \{t_e\} \leq t$
 at which $x(t_e)$ exceeds a given threshold $x t^{1 \over \mu}$ converges in
 distribution to a compound Poisson process with intensity
 and cluster probabilities that can be explicited \cite{emmbre,Sorknopoff}.

 \section{Characteristic function for $\mu >2$}
 
Within Renewal theory or Wiener-Hopf technique, 
the case $\mu >2$ does not play a special role and the previous results apply.
In the context of the characteristic
function used in \cite{Taka}, the case $\mu > 2$ can also be tackled by remarking that
the expression of the Laplace transform $\hat P(\beta)$ of a power law pdf $P$ with exponent $\mu$
presents a regular Taylor expansion in powers of $\beta$ up to the order
$l$ (where $l$ the integer part of
$\mu$) followed by a term of the form $\beta^{\mu}$. Let us give some details of this
derivation.
The Laplace transform
\begin{equation}
\hat P(\beta) \equiv \int_0^{\infty} dw P(w) e^{-\beta w} ,
\end{equation}
applied to (\ref{er}) yields
\begin{equation}
\hat P(\beta) =  C \int_1^{\infty} dw {e^{-\beta w} \over w^{1+\mu}} = 
\mu \beta^{\mu} \int_{\beta}^{\infty} dx {e^{-x} \over x^{1 + \mu}} .
\end{equation}
We have assumed, without loss of generality, that the power law holds for $x >1$.
Denote $l$ the integer part of $\mu$ ($l<\mu<l+1$). Integrating by part $l$
times, we get (for $C=\mu$) 
$$
\hat P(\beta) = e^{-\beta} \biggl(1 - {\beta \over \mu - 1} + ... + {(-1)^l
\beta^l \over (\mu - 1)(\mu - 2)...(\mu - l)} \biggl) + 
$$
\begin{equation}
+{(-1)^l \beta^{\mu} \over (\mu - 1)(\mu - 2)...(\mu - l)}
 \int_{\beta}^{\infty} dx e^{-x} x^{l - \mu} .
\end{equation}
This last integral is equal to
\begin{equation}
\beta^{\mu} \int_{\beta}^{\infty} dx e^{-x} x^{l - \mu} =
\Gamma(l+1-\mu) [\beta^{\mu} + \beta^{l+1} \gamma^*(l+1-\mu,\beta)] ,
\end{equation}
where $\Gamma$ is the Gamma function ($\Gamma(n+1)=n!$)
and 
\begin{equation}
\gamma^*(l+1-\mu,\beta)=
e^{-\beta} \sum_{n=0}^{+\infty} {\beta^n \over \Gamma(l+2-\mu+n)}
\end{equation}
is the incomplete Gamma function \cite{Abra}. We see that $\hat P(\beta)$ 
presents a regular Taylor expansion in powers of $\beta$ up to the order
$l$, followed by a term of the form $\beta^\mu$.
We can thus write
\begin{equation}
\hat P(\beta) = 1 + r_1 \beta + ..... + r_l \beta^l +
 r_\mu \beta^{\mu} + {\cal O}(\beta^{l+1}) ,
 \label{retfd}
\end{equation}
with $r_1 = -\langle x \rangle, \ r_2={\langle x^2 
\rangle \over 2}, ...$ are the moments of the powerlaw pdf
 and, reintroducing $C$, where $r_\mu$ is proportional
to the scale  parameter $C$. For small $\beta$, 
we exponentiate (\ref{retfd}) and rewrite $\hat P(\beta)$ under the form
\begin{equation}
\hat P(\beta) = \exp\left[\sum_{k=1}^l d_k \beta^k + d_\mu \beta^{\mu} \right]
,  \label{laplaform}
\end{equation}
where the coefficient $d_k$ can be simply expressed in terms of the $r_k$'s.
We recognize in this the transformation from the moments to the
cumulants. The expression (\ref{laplaform}) generalizes the canonical form
 of the characteristic function of the stable L\'evy laws, for
arbitrary values of $\mu$, and not solely for $\mu \leq 2$ for which they are
defined. The canonical form is recovered for $\mu \leq 2$ for which the coefficient
$d_2$ is not defined (the variance does not exist) and the only analytical
term is $\langle w \rangle \beta$ (for $\mu > 1$).
This rationalizes ``the lucky coincidence'' noticed by Takayasu et al.  \cite{Taka}
that the results obtained from the characteristic function were found to apply 
numerically for exponents $\mu >2$.

\section{Mechanism for the stretched exponential found by Takayasu et al.}

To mimick system size limitation, Takayasu et al. introduce a threshold $x_c$ 
such that for $|x(t)| > x_c$, $b(t) < 1$ and find a stretched exponential 
truncating the power law pdf beyond $x_c$. 
Frisch and Sornette \cite{Frisch}
have recently developed a theory of extreme deviations generalizing the central limit theorem
which, when applied to multiplication of random variables, predicts the generic
presence of stretched exponential pdf's. Let us briefly summarize the key ideas and
how it applies to the present context. First, we neglect $f(t)$ in (\ref{eq1}) for large
$x(t)$ ($x_c$ is supposed much larger than the characteristic scale of $f(t)$).
The problem thus boils down to determine the tail of the pdf for a product of random variables.

Consider the product
\begin{equation}
X_n = m_1 m_2 .... m_n
\label{productdef}
\end{equation}
If we denote $p(m)$ the pdf of the iid random variables $m_i$, then the pdf of $X_n$ is
\begin{equation}
P_n(X) \sim [p(X^{1 \over n})]^n  ,  \qquad
{\rm  for}\quad X \to \infty\quad {\rm and} \quad n \quad {\rm finite}  .
\label{devuuu}
\end{equation}
Equation (\ref{devuuu}) has a very intuitive interpretation\,:
the tail of $P_n(X)$ is controlled by the realizations where all terms in
the product are of the same order; therefore $P_n(X)$ is, to leading
order,  just the product
of the $n$ pdf's, each of their arguments being equal to the  common
value $X^{1 \over n}$.
When $p(x)$ is an exponential, a Gaussian or, more generally, of the
form $\propto \exp(-Cx^\gamma)$ with $\gamma>0$, then (\ref{devuuu})
leads to stretched exponentials for large $n$. For example, when $p(x)
\propto
\exp(-Cx^2)$, then $P_n(X)$ has a tail $\propto \exp(-CnX^{2/n})$.

Expression (\ref{devuuu}) is obtained directly by recurrence.
Starting from $X_{n+1} = X_n x_{n+1}$, we write the
equation for the pdf of $X_{n+1}$ in terms of the pdf's of $x_{n+1}$ and
$X_n$\,:
$$ P_{n+1}(X_{n+1}) = \int_0^{\infty} dX_n P_n(X_n) \int_0^{\infty}
dx_{n+1} p(x_{n+1}) \delta(X_{n+1} - X_n x_{n+1}) = $$
\begin{equation}
\int_0^{\infty} {dX_n \over X_n} P_n(X_n) p\left({X_{n+1} \over X_n}\right) ~.
\label{geneakfaf}
\end{equation}
The maximum of the integrand occurs for $X_n = (X_{n+1})^{n+1 \over
n}$ at which $X_n^{1 \over n} = {X_{n+1} \over X_n}$.  Assuming that
$P_n(X_n)$ is of the form (\ref{devuuu}), the formal application of
Laplace's method to (\ref{geneakfaf}) then directly gives that
$P_{n+1}(X_{n+1})$ is of the same form.
Thus, the property (\ref{devuuu}) holds for all $n$ to leading order
in $X$. See \cite{Frisch} for a more detailled derivation.

\section{Concluding remarks}

The process (\ref{eq1}) corresponds to a zero-dimensional
 process. An interesting extension consists in taking $x$ to be a function of 
 space ($d$-dimension) and time. Qualitatively, we thus get a
$d-$continuous
infinity of variables $x$, each of which follows a multiplicative
stochastic dynamics having the forms (\ref{eq1}) coupled to
nearest neighbors through a diffusion term.
Munoz and Hwa \cite{Hwa} find numerically a power law decay for the pdf of $x$ in the
$d$-dimensional case.

Autocatalytic equations lead to multiplicative
stochastic equations that are exactly tractable \cite{Graham} 
in the case of Gaussian multiplicative noise.
The process (\ref{eq1}) also describes accumulation
and discount in finance, perpetuities in insurance, ARCH processes in econometry, 
time evolution
of animal population with restocking \cite{Sorknopoff}. The
random map (\ref{eq1}) can also be applied to problems of population
dynamics, epidemics, investment portfolio growth, and immigration across
national borders \cite{Sorknopoff}.

\end{document}